\documentclass{mn2e}

\usepackage[dvips]{graphicx}

\begin{document}

\title[Galaxy satellites]
{Anisotropy in the distribution of satellites around primary galaxies in the 2DFGRS: the Holmberg effect}

\author[Sales \& Lambas]{Laura Sales, $^{1}$ and Diego G. Lambas,$^{1,2,3}$,
\\
$^1$ Grupo IATE, Observatorio Astron\'omico
de la Universidad Nacional de C\'ordoba,  Argentina.\\
$^2$ Consejo Nacional de Investigaciones Cient\'{\i}ficas
y T\'ecnicas.\\
$^3$ John Simon Guggenheim Fellow.
}

\maketitle

\begin{abstract}
We have analysed a sample of satellite and primary galaxies in the 2dF galaxy redshift survey. In our study we find a strong statistical evidence of the Holmberg effect (that is a tendency for satellites to avoid regions along the line defined by the primary plane) within 500 kpc of projected distance to the primary. This effect is present only when we restrict to objects with radial velocity relative to the primary $|\Delta v|<160$ km/s which correspond approximately to the mean of the distribution. 
We explore the dependence of this anisotropy on spectral type ($\eta$), colours, and luminosities of both primaries and satellites, finding that objects with a low present-day star formation rate present the  most significant effect.
\end{abstract}


\section{INTRODUCTION}

The origin and evolution of structures in the universe can be suitably understood within
 the hierarchical scenario.
In such schemes, present-day systems are the result of the aggregation of several sub-structures, and in particular, satellite galaxies can contribute significantly as building blocks of galaxy formation. 
However, present-day systems of satellites have evolved since their formation; and therefore the observations of such systems can provide invaluable information on the formation and evolution of galaxies.

There are several works that explore the characteristics of systems of satellites.
 Holmberg (1969) found that the distribution of galaxy satellites presents a lack of objects within $30^\circ$ of the primary disk plane ($\phi<30^\circ$).
 He searched for companions of nearby bright disk galaxies (the primaries) within 50 kpc of projected separation ($r_p$). 
His sample was composed of 218 satellites including both physical systems and interlopers that contaminated his photometrically selected sample.
 He found 45 objects within $\phi<30^\circ$ while 173 had $\phi>30^\circ$.
If the position angles relative to the primary plane are homogeneously distributed, then the number expected within $\phi<30^\circ$ would be $\sim$ 73.
After removing contamination by presumed interlopers (which are expected to be distributed isotropically) he concluded that there were no physical objects with $\phi<30^\circ$.    

The results remained controversial since Valtonen et al. (1978) found an excess of compact satellites with small angles relatives to the primary disk. Based on this work, Byrd et al. (1987) argued that the anisotropy of the Holmberg sample was mainly due to selection effects for he may have not considered these compact dwarf objects. The authors proposed a dynamical stripping of the gas content of the satellite as a possible explanation for the excess of such objects near the primary plane.

Some years later, the problem was investigated again in a statistical way by Zaritsky et al. (1997) using redshift data. 
They selected a sample of 115 satellites of 69 nearby primaries, these satellites have a maximum projected separation $r_p<500$ kpc. 
 These authors could not confirm the Holmberg effect in the inner 50 kpc since they had few objects at this distance , but they found a similar anisotropy effect with reliable statistical significance at larger separations, $r_p>250$ kpc.

One simple starting point to approach this problem is to assume 
an isotropic initial distribution of satellites and that evolution 
 generates the observed anisotropies.
 This scenario has been explored by Quinn $\&$ Goodman (1986), 
who found that anisotropy in the distribution of satellite positions
is not efficiently produced by dynamical evolution, even at small separation, $r_p<50$ kpc.

On the other hand, numerical simulations like those carried out by Abadi et al. (2002) indicate that, in a hierarchical scenario, the formation of 
 disks involve substantial accretion of satellites which strongly influence
 the angular momentum vector and therefore
 the primary plane.
 Thus, observed anisotropies in the distribution of the remaining satellites could be a direct consequence of this process.
In addition to this work, Pe\~narrubia et al. (2002) also based on numerical simulations, established that the time interval before the satellite disruption strongly depends 
 on the dark matter halo shape, the initial orbital eccentricities and inclinations,
 and on the mass fraction of satellites relative to the primary.

Our work aims to search for anisotropies of the angular distribution of satellites 
relative to the primary plane as well as exploring the possible
dependence on satellites properties. 
 Our analysis is based on the study of an unprecedentedly large statistical sample extracted from the 2dF galaxy redshift survey. 
In section 2 we present the data and selection criteria, in section 3 we show the main results, and in section 4 we summarise our conclusions.

\section{PRIMARY AND SATELLITE SAMPLES}

 The largest complete redshift survey available to date is the 2dFGRS which comprises information of objects on approximately 1500 square degrees of the sky in the north (NGC) and south (SGC) strips, and in several random fields.
This catalogue has been obtained with the multifibre technology at the Anglo-Australian Observatory, and provides photometric and spectroscopic data for 232155 galaxies with magnitudes  $b_j\le19.45$. 
It contains positions, magnitudes (in $b_j$ and R band), eccentricities, position angles, redshifts (well determined for 221414 galaxies) and the spectral classification through the $\eta$ parameter (which has a high confidence for objects with $z<0.1$).  

The selection criteria adopted in this work are the usual in the literature.
From 2dFGRS we have selected  1567 primary galaxies with $z<0.1$ restricted to absolute magnitude $B_j<-18$.
We have imposed an isolation criterion by  considering primaries with no companion galaxies within projected distance $r_p<700$ kpc and relative radial velocity difference $|\Delta v|<1000$ km/s, and with absolute magnitude difference $ B_j^{comp}-B_j^{prim}<1$. 
The value of the Hubble constant in this work is assumed to be $H_0=100h$ km/s $Mpc^{-1}$.

 After the selection of the primary sample, we searched for satellite galaxies around each primary. 
We define as satellites galaxies with $r_p<500$ kpc and  $|\Delta v|<500$ km/s
 which are at least 2 magnitudes fainter than their primary,
 $B_j^{prim}-B_j^{sat}>2$.
 With all these restriction our final sample consists of 3079 satellites.

A few primaries have too many satellites (up to $N=19$ objects).
 These objects are likely to be contaminated by group-like systems with the primary as 
the brightest group member. 
In order to exclude these spurious systems we have restricted the sample to 1498 primaries with $N<5$
 out of the 1567 original primaries.
 Furthermore, we notice that the removed systems show systematically high mean velocity dispersion,
  close to typical value of groups of galaxies, supporting our conservative cut-off in 
 the number of members per system. 

We have also imposed an additional restriction to the eccentricities of the primaries in order to deal with well determined relative angles $\phi$ of satellite position with respect to the primary plane projected into the first quadrant.
 Thus, we have not considered primaries with $e\le0.1$

\section{RESULTS}

The sample selected under the criteria described in the previous section allows us to investigate the satellite position angle distribution relative to the primary plane disk ($\phi$). 

\begin{figure}
\includegraphics[width=84mm]{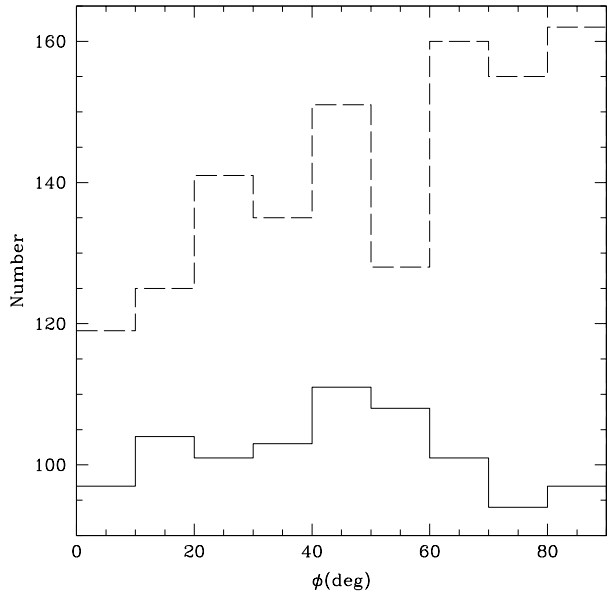}
\caption{Distribution of position angles of satellites for two subsamples corresponding to high and low relative velocity to the primary.
 The solid line represents satellites with $|\Delta v|>160$ km/s and the dotted line satellites with $|\Delta v|<160$ km/s.
 The adopted threshold $|\Delta v|= 160$ km/s is approximately the mean value of the satellite relative velocity distribution (163.9 km/s).} 
\label{vel}
\end{figure}

 We have divided the sample into objects with high or low $|\Delta v|$ compared to $|\Delta v|=160$km/s (a value close to  the mean $|\Delta v|=163.9$ km/s).
 The distribution of satellite position angles for these two subsamples are shown in Fig. \ref{vel}.
  From this figure one can notice the deficiency of satellites with low $\phi$ values in 
the subsample of satellites with relative velocity $|\Delta|<160$ km/s.
 On the contrary, the high velocity subsample is roughly isotropic.

We may quantify the departure from isotropy defining the ratio $f$ of objects in planar ($\phi<30^\circ$) to polar ($\phi>60^\circ$) angular positions $f=N_{<30}/N_{>60}$,
 for isotropy $f=1$.  
The resulting values are:
$f=0.80+-0.04$ for $|\Delta v|<160$ km/s and
$f=0.98+-0.08$ for $|\Delta v|>160$ km/s. 
 Quoted uncertainties in this work were calculated  based on 20 bootstrap re-samplings of the data.

\begin{figure}
\includegraphics[width=84mm]{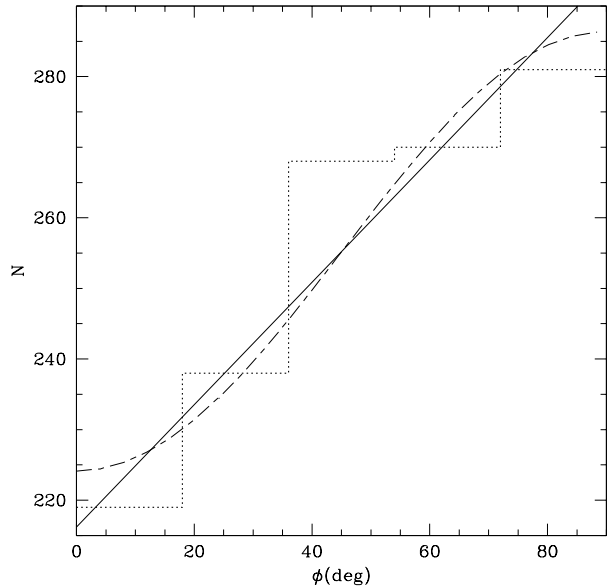}
\caption{Distribution of satellite position angles for subsample with $|\Delta v|<160$ km/s. We show two possible fit to the data. In solid curve is the linear relation and in long-dashed ones is the double cosine function. }
\label{comp}
\end{figure}

 Other useful statistical measures of this anisotropy effect can be obtained by fitting a function to the $\phi$ distribution.
 We have adopted a double cosine function 
$N/\bar{N}=Acos(2\phi)+B$, and a linear fit  $N/\bar{N}=a_{lin}\phi+b$
where $\bar{N}$ is the mean number of objects and  A and $a_{lin}$ 
are the anisotropy parameter in each fit. 
 For an isotropic distribution on $\phi$ we expect $A=a_{lin}=0$.
 Either function, the double cosine and the linear relation, are suitable to fit the data,  see for instance Fig. \ref{comp}. 
We notice, however that
the double cosine function is less sensitive to noise fluctuations at the extremes 
due to poor number statistics and therefore might provide  more conservative estimate of the 
deviation from isotropy. For this reason we will mainly refer to the
A parameter although both A and $a_{lin}$ values are listed in the tables corresponding to the different 
subsamples analysed.

\begin{table}
\center
\caption{Anisotropy parameters $a_{lin}$ and A, and their bootstrap derived error estimate.}
\begin{tabular}{|c c c c c c c}
Sample & Characteristics &  N   & $a_{lin}$& $\sigma_a$ & A   & $\sigma_A$ \\
$S-vel_<$ & $|\Delta v|<160$ km/s &0.4& 0.1& 1276 & 0.12 & 0.04 \\
$S-vel_>$ & $|\Delta v|>160$ km/s &0.0& 0.1& 916 & 0.01 & 0.04 \\\\
$S-rp_<$ & $|\Delta v|<160$ km/s,\\
         & $r_p<250$ kpc $h^{-1}$ &0.2& 0.1& 713 & 0.11 & 0.05\\     
$S-rp_>$ & $|\Delta v|<160$ km/s,\\
         & $r_p>250$ kpc $h^{-1}$ &0.2& 0.1& 691 & 0.12 & 0.06\\     

\end{tabular}
\label{tvel}
\end{table}

 In table \ref{tvel} we list the number of satellites in each subsample ($|\Delta v|<160$ km/s and $|\Delta v|>160$ km/s) and  
 the corresponding $a_{lin}$ and A values. 
 The A parameter for subsample $S-vel_<$ departs from isotropy by more than $\sim3\sigma$ which contrasts with the behaviour of subsample $S-vel_>$ where $A\simeq0$.
 This reflects the trends shown in Fig. \ref{vel}, and is consistent with our previous estimate of $f$.
The fact that the anisotropy signal is not significant in the high relative velocity
subsample ($|\Delta v|>160$ km/s)
 suggests a higher contamination due to interlopers for these objects (see for instance Zaritsky et al. 1993).

 The number of objects per bin is $N=N_s(\phi)+N_i(\phi)$, where $N_s$ and $N_i$ are the number of physical satellites and interlopers as function of angle, respectively.
 False satellites should have a homogeneous distribution of $\phi$ angles, so
 we take $N_i(\phi)=N_i$.
 On the other hand, the distribution of real satellites respect to the primary plane may present anisotropies. The observed anisotropy parameter $A_{obs}$  is related to the real anisotropy parameter $A_{real}$ by: $A_{obs}=(<N_s>/<N_s>+N_i) A_{real}$.
 Thus, even a modest observed anisotropy $\simeq10\%$, can imply a substantially large real anisotropy.

As suggested by Zaritsky (1992), the relative fraction $P/T$ of objects with $\Delta v>0$ for
a given sample can provide a useful measure of the degree of contamination given the different volumes involved. 
If a sample is seriously contaminated by interlopers, we should obtain $P/T >> 0.5$.

Therefore, we calculated $P/T$ for two subsamples, satellites with $|\Delta v|<160$ km/s and satellites with $|\Delta v|>160$ km/s. 
 For these two subsamples $P/T$ values are $0.54\pm0.02$ and $0.51\pm0.03$ respectively, 
 so that from this particular test we cannot argue for substantial contamination in the high velocity subsample. Thus the different behaviour of high and low relative velocity satellites could be a real effect.

 In his pioneer work, Holmberg found an excess of polar satellites within 50 kpc from the primary; nevertheless,  Zaritsky et al. (1996) could not reproduce this result, but they found such an excess at larger separations (250 kpc$<r_p<$500 kpc) from the primaries.
 We have searched for possible dependences of the detected anisotropies on projected separations to the primaries.
 The resulting anisotropy parameter A are displayed in table \ref{tvel}, where it can be appreciated the lack of strong dependence on projected separation.
 From now on, we restrict our sample of satellites to those objects with $|\Delta v|<160$ km/s.
 
\begin{figure}
\includegraphics[width=84mm]{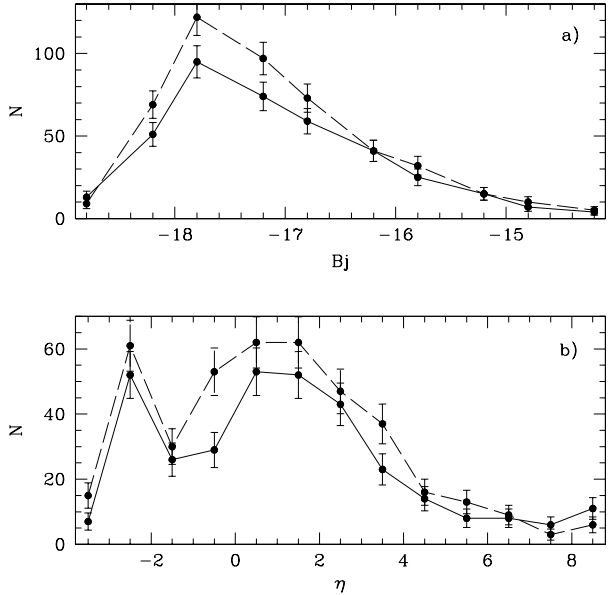}
\caption{Different satellite parameter distributions. Dashed (solid) curves are polar (coplanar) angular positions. In upper panel is $B_j$ distribution, in the bottom is represented the spectral type $\eta$ of the satellites.
 The error bars correspond to Poisson uncertainties. }
\label{fig2}
\end{figure}

 In order to investigate if there are any preferred satellite characteristics which show a more significant anisotropy signal, we have analysed the distributions of absolute magnitude and spectral type of satellites in polar ($\phi>60^\circ$) and planar ($\phi<30^\circ$) positions.
 The objects with $\phi<30^\circ$ and $\phi>60^\circ$ are 385 and 477 respectively. 
We show in 
Fig \ref{fig2} $a)$ and $b)$,  the corresponding distributions of absolute magnitude $B_j$ and spectral type parameter $\eta$ for these two cases.
 The error bars correspond to Poisson uncertainties.
 It can be appreciated in these figures that the excess of objects with $\phi>60^\circ$,
 is distributed quite uniformly along the explored range of luminosities and spectral types.

In order to quantify the anisotropy for these subsamples we show in table \ref{tsat} the corresponding values of $a_{lin}$ and A parameters. 

\begin{table}
\center
\caption{The anisotropy parameters $a_{lin}$ and A for samples of different satellites properties.}
\begin{tabular}{|c c c c c c c}
Sample & Characteristics &  N &$a_{lin}$&$\sigma_a$&  A   & $\sigma_A$ \\
S1 & $B_j^{sat}<-17.3$ & 633 & 0.3 & 0.1 & 0.15 & 0.04 \\
S2 & $B_j^{sat}>-17.3$ & 643 & 0.1 & 0.1 & 0.09 & 0.04 \\\\
S3 & $\eta^s<1.1$ & 580 & 0.3 & 0.1 & 0.16 & 0.04 \\
S4& $\eta^s>1.1$ & 643 & 0.1 & 0.1 & 0.08 & 0.04\\\\ 
S5 & $(b_j-R)^s<0.83$ & 644 & 0.3 & 0.09 & 0.15 & 0.05 \\
S6 & $(b_j-R)^s>0.83$ & 632 & 0.2 & 0.09 & 0.09 & 0.06 \\
\end{tabular}
\label{tsat}
\end{table}
 
The adopted thresholds in absolute magnitude $B_j$, spectral type $\eta$ and colour index $b_j-R$ were chosen in order to divide the samples into approximately equal number subsets. 
 
By inspection to table 2 it can be seen that the anisotropy of sample S2, although 
less statistically significant that in sample S1, still shows an excess of polar satellites at a $\sim2.5\sigma$ level. Thus, this 
result does not support the hypothesis of Byrd et al. (1987) where 
the lack of satellites with $\phi<30^\circ$ in Holmberg's data
would correspond to a combination of dynamical stripping of gas
and selection effects, where faint compact satellites should be found preferentially near the primary plane. 

The results for subsample S3 indicate that the satellites with low star formation activity  present the strongest anisotropy signal compared to those of high star formation activity (subsample S4).
Subsamples S5 and S6 indicate that the satellites characterized by low colour index values show a stronger excess near the primary poles.
 This can be appreciated from Fig \ref{fig2}$b)$, where is evident that the intermediate spectral type satellites ($-1.4<\eta<1.1$) show a strong deficiency at $\phi<30^\circ$.

We have also explored for possible dependencies of the anisotropy on properties of the primaries, namely luminosity, spectral type and colour index.
In table \ref{tpri} we list N, $a_{lin}$, $\sigma_a$, A and $\sigma_A$ for subsamples selected according 
to primary properties (regardless of satellites characteristics).

\begin{table}
\center
\caption{The anisotropy parameters $a_{lin}$ and A for samples of different primary properties.}
\begin{tabular}{|c c c c c c c}
Sample & Characteristics &  N  & $a_{lin}$ & $\sigma_a$ &  A   & $\sigma_A$ \\
P1 & $B_j^{prim}<-20$ & 630 & 0.2 & 0.1 & 0.10 & 0.07 \\
P2 & $B_j^{prim}>-20$ & 643 & 0.3 & 0.1 & 0.14 & 0.05 \\\\
P3 & $\eta^p<-1.4$ & 686 & 0.4 & 0.1 & 0.18 & 0.03 \\
P4 & $\eta^p>-1.4$ & 581 & 0.1 & 0.1 & 0.03 & 0.06\\\\ 
P5 & $(b_j-R)^p<1.13$ & 645 & 0.1 & 0.1 & 0.08 & 0.05 \\
P6 & $(b_j-R)^p>1.13$ & 631 & 0.3 & 0.1 & 0.16 & 0.05 \\
\end{tabular}
\label{tpri}
\end{table}

\begin{figure}
\includegraphics[width=84mm]{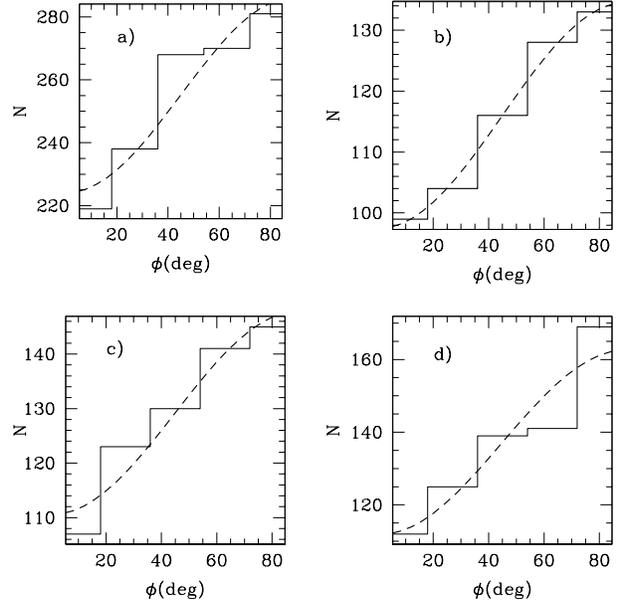}
\caption{Double cosine fits for different subsamples: $a)$ $S-vel_<$, $b)$ S3, $c)$ P2 and $d)$ P3.}
\label{fig3}
\end{figure}

Different ranges of absolute magnitudes of the primaries show different $\phi$ distributions (see first two lines in table \ref{tpri}).
 For faint primaries (P2) the angular positions of satellites are well reproduced by our double cosine fit, where the anisotropy parameter differs from zero by $\simeq3\sigma$.
However, for bright primaries (P1) this effect is not statistical significant.    
In a recent paper, Prada et al. (2003) have shown that the luminosity and halo mass of a galaxy are closely related.
 This results suggest that the anisotropy of the satellite distribution
would not have a tidal origin, since in this case we would expect satellites of bright primaries to show the strongest anisotropy, contrary to our findings.
   
 On the other hand, we detect a tight dependence of the satellite anisotropy distribution on the spectral type of the primary.
In table \ref{tpri} it can be seen that satellites of passive star forming primaries (P3)
 show a departure from isotropy of at least $\sim6\sigma$. 
 A different behavior is found for the sample of primaries with significant star formation activity (P4) which show a remarkable isotropic distribution of satellites.
Regarding the colour indexes of primaries we find a more significant effect for primaries with larger values of $b_j-R$ colour index (sample P6).

In Fig. \ref{fig3} we show four different subsamples with the high anisotropy signal:  $a)$ satellites with $|\Delta v|<160$ km/s ($S-vel_<$), $b)$ satellites with $\eta^s<1.1$ (S3), $c)$ and $d)$ primaries with absolute magnitude $B_j>-20$ (P2) and $\eta^p<-1.4$ (P3) respectively.

 A useful test to check the real presence of the signal in our analysis is to restrict 
 to primaries with high eccentricity values ($e$).
 We have performed this test for subsample P3 (primaries with low star formation activity) which show a large anisotropy effect.
 In table \ref{tem} we show the results obtained for this sample for different threshold values of $e$.
 It can be seen that the strongest signal corresponds to primaries which are nearly edge-on ($e>0.4$) which gives substantial confidence to our findings.

\begin{table}
\center
\caption{The anisotropy parameters $a_{lin}$ and A for samples of passively star forming primaries ($\eta^p<-1.4$) with different eccentricities.}
\begin{tabular}{|c c c c c c c }
Sample & Characteristics &  N   & $a_{lin}$ & $\sigma_a$ &  A   & $\sigma_A$2 \\
P3 & $e>0.1$ & 686 & 0.4 & 0.1 & 0.18 & 0.03 \\
P3$_a$ & $0.1<e<0.4$ & 338 & 0.2 & 0.1 & 0.15 & 0.08 \\
P3$_b$ & $e>0.4$ & 337 & 0.3 & 0.1 & 0.26 & 0.08 \\
\end{tabular}
\label{tem}
\end{table}

\section{SUMMARY}

We have carried out an analysis of the anisotropy of satellite angular positions with respect to the primary plane.
 The sample was selected in redshift space from the 2dFGRS and is large enough for studying possible dependencies of anisotropies on both primary and satellite properties. 

We confirm the existence of the effect first claimed by Holmberg, that is, the preference of the satellites to lie near the pole of the primary.
 We note however that this effect is present up to relative projected separations $r_p<500$ kpc. 

We use an anisotropy parameter A suitable to measure the departure from isotropy in the different subsamples analysed.
 Nevertheless, this parameter indicates a lower limit to the real anisotropy, given the contamination by false satellites expected to be uniformly distributed with respect to the primary plane.
 
The deviation from isotropy is detected at least at $\sim3\sigma$ level for objects with low values of relative velocity $|\Delta v|<160$ km/s while the high velocity subsample is consistent with isotropy.
Although in principle this effect could be related to different levels of contamination by interlopers, Zaritsky 1992 test provides comparable values of contamination in both high and low relative velocity subsamples.

 By inspection to the luminosity of the primaries, we conclude that the anisotropy of the distribution of satellites around faint primaries shows a high statistical significance ($\simeq3\sigma$). 
 
 The most important signal was detected in the sample of satellites corresponding to passive star forming primaries, where the distribution of $\phi$ angles departs from isotropy at the $\simeq6\sigma$ level; 
 while the opposite behaviour is present in primaries with $\eta>-1.4$, for which the A parameter results consistent with zero.
 Moreover, we also notice that the signal is even higher when we raise the eccentricity threshold of the primaries.
 We also have considered the different spectral types of the satellites, finding the strongest anisotropy for satellites with $\eta<1.1$ (objects with little present day star formation activity). 
 We notice that primaries with a low value of $\eta$ in our sample also have satellites with poor star formation activity.      

 The isolation criteria imposed to our sample lowers the probability of including a significant number of elliptical galaxies.
 Thus, most primaries in our sample are expected to be of late type morphology.
 Due to the morphology density relation we expect only 10$\%$ of our isolated sample of primaries to be ellipticals (Whitmore et al. 1993). 

We could in principle interpret the Holmberg effect as being mainly caused by evolutionary processes.
 Systems dominated by old stellar population objects have formed early, so the physical processes that may have removed  satellites near the primary plane have been present for a large time.
 By contrast, in younger primary and satellite systems, we would expect a larger fraction of objects that have not been accreted yet on to the primary so that they would show an isotropic distribution of satellites.
However, as discussed by Quinn $\&$ Goodman (1986), it may be difficult to produce anisotropies in the distribution of satellites from pure dynamical evolution, specially at large scale  $r_p<500$ kpc.
 More likely, we suggest that according to Abadi et al. (2002) results, such anisotropies could be the remnants of the formation of the disks of the primaries, where substantial orbital momentum is transfered into primary rotation during the early stages of galaxy formation. Thus, the anisotropy effects we detect at large projected separations, could be the relics of such mechanism.

We also argue that the significant dependence of the anisotropy effect on the relative velocity between the satellite and the primary could be an important point of disagreement between the results of previous works.

\section*{Acknowledgments}
We thank the Referee for helpful comments and suggestions which greatly improved the previous version of this paper. This work was partially supported by the
 Consejo Nacional de Investigaciones Cient\'{\i}ficas y T\'ecnicas,
Agencia de Promoci\'on de Ciencia y Tecnolog\'{\i}a,  Fundaci\'on Antorchas
 and Secretar\'{\i}a de Ciencia y
T\'ecnica de la Universidad Nacional de C\'ordoba.

\end{document}